\documentclass[prd,aps,nofootinbib,notitlepage,showpacs,preprintnumbers]
{revtex4-1}
\usepackage{graphicx,epsf,amsmath,amsfonts,amssymb,amsbsy}
\usepackage{epsfig}
\newcommand{\ds}{\displaystyle}
\newcommand{\vev}[1]{\langle#1\rangle}
\newcommand{\mat}{\left ( \begin{array}}
\newcommand{\emat}{\end{array} \right )}
\newcommand{\vect}{\left ( \begin{array}{c}}
\newcommand{\evect}{\end{array} \right )}

\begin{document}


\title{Superconductivity phenomenon induced by external in-plane
  magnetic field in (2+1)-dimensional Gross--Neveu type model}
\author{K.G. Klimenko $^{a,b}$,
R.N. Zhokhov $^{a}$ and V.Ch. Zhukovsky $^{c}$}
\affiliation{$^{a}$ Institute for High Energy Physics,
142281, Protvino, Moscow Region, Russia}
\affiliation{$^{b}$ University "Dubna" (Protvino branch),
142281, Protvino, Moscow Region, Russia}
\affiliation{$^{c}$ Faculty of
Physics, Moscow State University, 119991, Moscow, Russia}

\begin{abstract}
Phase structure of the (2+1)-dimensional model with four-fermion
interaction of spin-1/2 quasiparticles (electrons) both in the
fermion-antifermion (or chiral) and fermion-fermion (or
superconducting) channels is considered at nonzero chemical
potential $\mu$ and under the influence of an in-plane, i.e. parallel to a
system sheet, external magnetic field $\vec B_\parallel$. It is
shown that at sufficiently large values of $\mu$ and/or $\vec
B_\parallel$ the Cooper pairing (or superconducting) phase appears
in the system at arbitrary relation between coupling constants,
provided that there is an (arbitrary small) attractive interaction
in the superconducting channel. In particular, at sufficiently weak
attractive interaction in the chiral channel, the Cooper pairing
occurs even at infinitesimal values of $\mu$ and/or $\vec
B_\parallel$.
The superconducting phase of the model is always a paramagnetic one.
\end{abstract}



\maketitle

\section{ Introduction}

Recently much attention has been paid to investigation of
(2+1)-dimensional quantum field theories (QFT) and, in particular,
to models with four-fermion interactions of the Gross--Neveu (GN)
\cite{GN} type. Partially,  this interest is explained by more
simple  structure of QFT in two-, rather than in three spatial
dimensions. As a result, it is much easier to investigate
qualitatively such real physical phenomena as dynamical symmetry
breaking \cite{GN,semenoff1,semenoff2,rosenstein,klimenko2,klimenko3,
inagaki,kanemura,hands,hands21,hands22,bashir,bashir2,khanna} and color superconductivity \cite{toki,kohyama1,kohyama2}
as well as to model phase diagrams
of real quantum chromodynamics \cite{kneur1,kneur2} etc. in the
framework of (2+1)-dimensional QFT. Another example of this kind
is the spontaneous chiral symmetry breaking induced by external magnetic
fields. This effect was for the first time studied also in terms of
(2+1)-dimensional GN models \cite{klimenko11,klimenko12,klimenko13,Gusynin,klimenk}. Moreover, these
theories are very useful in developing new QFT techniques like the
optimized perturbation theory \cite{kneur1,kneur2,k1,k2}, and so on.

However, there is yet another more serious motivation for studying
(2+1)-dimensional  QFT. It is supported by the fact that there are
many condensed matter systems which, firstly, have a (quasi-)planar
structure and, secondly, their excitation spectrum is described
adequately by relativistic Dirac-like equation rather than by
Schr\"{o}dinger one. Among these systems are the high-T$_c$ cuprate
and iron  superconductors \cite{davydov1,davydov2}, the one-atom thick layer
of carbon atoms, or graphene, \cite{niemi,castroneto} etc. Thus,
many properties of such condensed matter systems can be explained in
the framework of various (2+1)-dimensional QFT, including the
GN-type models (see, e.g., \cite{Semenoff:1998bk,babaev1,babaev2,Fialkovsky,caldas,zkke,zkke2,zkke31,zkke32,ohsaku,marino,marino3,marino2,zhokhov} and references therein).

In the recent papers \cite{marino,marino3,marino2,zhokhov} a competition between chiral symmetry  breaking (excitonic pairing) and superconductivity phenomenon (Cooper pairing) was investigated in the framework of (2+1)-dimensional GN-type models. There the influence of such external factors, as temperature $T$, chemical potential $\mu$ and
external magnetic field $\vec B_\perp$ perpendicular to the system plane, on the chiral and electromagnetic $U(1)$ symmetries was
studied. In particular, it was shown in \cite{marino2} that
sufficiently strong perpendicular  magnetic field $\vec B_\perp$
destroys the superconducting state of a planar system.

In the present paper the (2+1)-dimensional GN-type model, which
describes four-fermion  interaction of quasiparticles with spin
$1/2$ (electrons) both in the chiral (with coupling constant $G_1$)
and Cooper pairing (with coupling constant $G_2$) channels, is
considered at $T=0$ and $\mu\ne 0$. In addition, we suppose that the
planar system of electrons is subjected to an external magnetic
field $\vec B=\vec B_\perp+\vec B_\parallel$ in such a way that
$\vec B_\perp=0$ ($\vec B_\parallel$ is the in-plane, i.e. parallel
to the system plane, magnetic field). As a result, in our
consideration the external  magnetic field $\vec B$ couples only to
the spin of electrons and not to its orbital angular momentum. (The
interaction between $\vec B$ and orbital angular momentum of
electrons appears in planar systems only at $\vec B_\perp\ne 0$.) In
this case the way to account for the influence of external parallel
magnetic field on the phase structure of any planar system is to
introduce the Zeeman terms into Lagrangian, which are really due to
Zeeman effect. These terms modify effectively the chemical
potentials corresponding to the electrons with different spin
projections along the direction of an external magnetic field.
\footnote{In the recent paper \cite{caldas} the magnetization of
planar systems exposed to an external in-plane magnetic field was
investigated in the framework of the same (2+1)-dimensional GN model
but in the particular case $G_1\ne 0$, $G_2=0$. In particular, it
was shown there that at sufficiently high values of $\vec
B_\parallel$ there is a restoration of the chiral symmetry of the
model.} In doing so we show that the external magnetic field $\vec
B_\parallel$ parallel to the system plane induces the
superconductivity phenomenon in the initial model even at
infinitesimal coupling $G_2$ of the Cooper pairing channel. If in
addition the coupling constant $G_1$ is sufficiently small, then
superconductivity appears at arbitrary weak in-plane magnetic field
$\vec B_\parallel$ (of course, provided that $G_2>0$).

The paper is organized as follows. In Sec. II the (2+1)-dimensional
GN-type model  with four-fermion interactions in the
fermion-antifermion (or chiral) and fermion-fermion (or
superconducting) channels is presented. \footnote{It is a
  (2+1)-dimensional generalization, made in \cite{zhokhov}, of the
  initially (1+1)-dimensional well-known model by Chodos et al
  \cite{chodos}. }
Here the unrenormalized
thermodynamic potential (TDP) of the model is obtained in the leading order of the large $N$ technique, taking into account the nonzero values of the
chemical potential $\mu$ and the external in-plane magnetic field
$\vec B_\parallel$ (temperature is put equal to zero). In the next Sec.
III a renormalization group invariant expression for the TDP in the leading order of $1/N$ expansion is obtained. The TDP global minimum point provides us with chiral and Cooper pairing condensates. In Sec. IV A phase structure of the model is described at $\mu=0$ and $\vec B_\parallel = 0$.
Finally, in Sec. IV B the $(\mu,|\vec B_\parallel|)$-phase
diagrams as well as the behavior of gaps and different thermodynamic quantities of the system such as particle density, magnetization and magnetic susceptibility are presented for some representative values of coupling constants. We show in this section that for arbitrary relations
between coupling constants superconductivity is induced in the
system at sufficiently large values of $\mu$ and/or in-plane
magnetic field $\vec B_\parallel$. In particular,
it is found out that infinitesimal values of $\vec
B_\parallel$ induce the superconductivity phenomenon in the case of
a rather weak attractive interaction in the fermion-antifermion
channel.

\section{ The model and its thermodynamic potential}
\label{effaction}

We investigate the influence of an external magnetic field $\vec B$
on the phase structure of (2+1)-dimensional version of the Chodos et
al. model \cite{zhokhov,chodos} which describes low-energy dynamics
of quasiparticles (electrons) both in the fermion-antifermion (or
chiral) and fermion-fermion (or Cooper pairing) channels. In
addition, we take into account the fact that there are two spin
projections, $\pm$1/2, of electrons on the direction of the magnetic
field $\vec B$. If external magnetic field is  parallel to the
system plane, i.e. $\vec B=\vec B_\parallel$, then  the Lagrangian
has the following form
\begin{eqnarray}
 L=\sum_{k=1}^2\bar \psi_{ka}\Big [\gamma^\rho i\partial_\rho
+\mu\gamma^0-\nu(-1)^k\gamma^0\Big ]\psi_{ka}+ \frac{G_1}{N}\left
(\sum_{k=1}^{2}\bar \psi_{ka}\psi_{ka}\right )^2+\frac{G_2}{N}\left
(\sum_{k=1}^{2} \psi_{ka}^T C\psi_{ka}\right )\left
(\sum_{j=1}^{2}\bar \psi_{jb}C\bar\psi_{jb}^T\right ), \label{1}
\end{eqnarray}
where the summation over the repeated indices $a,b=1,...,N$ of the
internal $O(N)$ group as well as repeated Lorentz indices
$\rho=0,1,2$ is implied. For each fixed values of $k=1,2$ and
$a=1,...,N$ the quantity $\psi_{ka}(x)$ in (1) means the massless
Dirac fermion field, transforming over a reducible 4-component
spinor representation of the (2+1)-dimensional Lorentz group.
Moreover, all these Dirac fields $\psi_{ka}(x)$ are composed into
two fundamental multiplets, $\psi_{1a}(x)$ and $\psi_{2a}(x)$
($a=1,...,N$), of the internal auxiliary $O(N)$ group, which is
introduced here in order to make it possible to perform all the
calculations in the framework of the nonperturbative large $N$
expansion method. We suppose that spinor fields $\psi_{1a}(x)$ and
$\psi_{2a}(x)$ ($a=1,...,N$) correspond to electrons with spin
projections 1/2 and -1/2 on the direction of external magnetic
field, respectively. In (1) the symbol $T$ denotes the transposition
operation, $\mu$ is a fermion number chemical potential and the
$\nu$-term is introduced in order to take into account the Zeeman
interaction energy of electrons with external magnetic field $\vec
B_\parallel$. Hence, in our case $\nu=g\mu_B B/2$, where $B=|\vec
B_\parallel|$, $g$ is the spectroscopic Lande factor (in what
follows  it is supposed throughout the paper that $g=2$) and $\mu_B$
is the Bohr magneton. Moreover, $C\equiv\gamma^2$ is the charge
conjugation matrix. The algebra of the $\gamma^\rho$-matrices as
well as their particular representations are given, e.g., in
\cite{zhokhov}. The model (1) is invariant under the discrete chiral
transformation, $\psi_{ka}\to\gamma^5\psi_{ka}$ (the particular
realization of the $\gamma^5$-matrix is also presented in
\cite{zhokhov}), as well as with respect to the transformations from
the continuous $U(1)$ fermion number group, $\psi_{ka}\to\exp
(i\alpha)\psi_{ka}$ ($k=1,2$, $a=1,...,N$), responsible for the
fermion number conservation or, equivalently, for the electric
charge conservation law in the system under consideration.
Certainly, there is $O(N)$ invariance of the Lagrangian (1).

The linearized version of Lagrangian (\ref{1}) that
contains auxiliary bosonic fields $\sigma (x)$, $\Delta(x)$,
$\Delta^{*}(x)$ has the form
\begin{eqnarray}
{\cal L}\ds =
 -\frac{N\sigma^2}{4G_1} -\frac{N\Delta^{*}\Delta}{4G_2}+\sum_{k=1}^2\left
 [\bar\psi_{ka}\Big (\gamma^\rho i\partial_\rho
+\mu_k\gamma^0 -\sigma \Big )\psi_{ka}-
 \frac{\Delta^{*}}{2}\psi_{ka}^TC\psi_{ka}
-\frac{\Delta}{2}\bar\psi_{ka} C\bar\psi_{ka}^T\right ], \label{2}
\end{eqnarray}
where $\mu_1=\mu+\nu$, $\mu_2=\mu-\nu$ and from now on $\nu=\mu_B B$
(in this formula and below the summation over repeated indices is
implied). Clearly, the Lagrangians (\ref{1}) and (\ref{2}) are
equivalent, as can be seen by using the Euler-Lagrange equations of
motion for scalar bosonic fields which take the form
\begin{eqnarray}
\sigma (x)=-2\frac{G_1}{N}\sum_{k=1}^2\bar\psi_{ka}\psi_{ka},~~
\Delta(x)=-2\frac{G_2}{N}\sum_{k=1}^2\psi_{ka}^TC\psi_{ka},~~
\Delta^{*}(x)=-2\frac{G_2}{N}\sum_{k=1}^2\bar\psi_{ka}
C\bar\psi_{ka}^T. \label{3}
\end{eqnarray}
One can easily see from (\ref{3}) that the neutral field $\sigma(x)$
is a real quantity, i.e. $(\sigma(x))^\dagger=\sigma(x)$ (the
superscript symbol $\dagger$ denotes the Hermitian conjugation), but
the (charged) difermion fields $\Delta(x)$ and $\Delta^*(x)$ are
mutually Hermitian conjugated complex quantities, so
$(\Delta(x))^\dagger= \Delta^{*}(x)$ and vice versa. If the
difermion field $\Delta(x)$ has a nonzero ground state expectation
value, i.e. $\vev{\Delta(x)}\ne 0$, the Abelian fermion number
$U(1)$ symmetry of the model is spontaneously broken down and the
superconducting phase is realized in the model.
(Note, at $T=0$ a continuous symmetry breaking is allowed to occur
in two spatial dimensions. The clarifying discussion is presented,
e.g., in the papers \cite{rosenstein,zhokhov}).
However, if
$\vev{\sigma (x)}\ne 0$ then the discrete chiral symmetry of the
model is spontaneously broken.

Let us now study the phase structure of the four-fermion model (1)
by starting from the equivalent semi-bosonized Lagrangian (\ref{2}).
In the leading order of the large $N$ approximation, the effective
action ${\cal S}_{\rm {eff}}(\sigma,\Delta,\Delta^{*})$ of the
considered model is expressed by means of the path integral over
fermion fields
$$
\exp(i {\cal S}_{\rm {eff}}(\sigma,\Delta,\Delta^{*}))=
  \int\prod_{k=1}^{2}\prod_{a=1}^{N}[d\bar\psi_{ka}][d\psi_{ka}]\exp\Bigl(i\int {\cal
  L}\,d^3 x\Bigr),
$$
where
\begin{eqnarray}
&&{\cal S}_{\rm {eff}} (\sigma,\Delta,\Delta^{*}) =-\int d^3x\left
[\frac{N}{4G_1}\sigma^2(x)+ \frac{N}{4G_2}\Delta
(x)\Delta^{*}(x)\right ]+ \widetilde {\cal S}_{\rm {eff}}. \label{5}
\end{eqnarray}
The fermion contribution to the effective action, i.e.\  the term
$\widetilde {\cal S}_{\rm {eff}}$ in (\ref{5}), is given by
\begin{eqnarray}
\exp(i\widetilde {\cal S}_{\rm
{eff}})&=&\int\prod_{l=1}^{2}\prod_{a=1}^{N}[d\bar\psi_{la}][d\psi_{la}]\exp\Bigl\{
i\int\sum_{k=1}^2\Big [\bar\psi_{ka}\Big (\gamma^\rho i\partial_\rho
+\mu_k\gamma^0 -\sigma \Big )\psi_{ka}\nonumber\\&&~~~~~~~~~~~~~
-\frac{\Delta^{*}}{2}\psi_{ka}^TC\psi_{ka}
-\frac{\Delta}{2}\bar\psi_{ka} C\bar\psi_{a}k^T\Big ] d^3 x\Bigr\}.
\label{6}
\end{eqnarray}
The ground state expectation values $\vev{\sigma(x)}$,
$\vev{\Delta(x)}$, and $\vev{\Delta^*(x)}$ of the composite bosonic
fields are determined by the saddle point equations,
\begin{eqnarray}
\frac{\delta {\cal S}_{\rm {eff}}}{\delta\sigma (x)}=0,~~~~~
\frac{\delta {\cal S}_{\rm {eff}}}{\delta\Delta (x)}=0,~~~~~
\frac{\delta {\cal S}_{\rm {eff}}}{\delta\Delta^* (x)}=0. \label{7}
\end{eqnarray}
For simplicity, throughout the paper we suppose that the above
mentioned ground state expectation values do not depend on
spacetime coordinates, i.e.
\begin{eqnarray}
\vev{\sigma(x)}\equiv M,~~~\vev{\Delta(x)}\equiv \Delta,~~~
\vev{\Delta^*(x)}\equiv \Delta^*, \label{8}
\end{eqnarray}
where $M,\Delta,\Delta^*$ are constant quantities. In fact, they are
coordinates of the global minimum point of the thermodynamic
potential (TDP) $\Omega (M,\Delta,\Delta^*)$. In the leading order
of the large $N$ expansion this quantity is defined by the following
expression:
\begin{equation*}
\int d^3x \Omega (M,\Delta,\Delta^*)=-\frac 1N{\cal S}_{\rm
{eff}}\{\sigma(x),\Delta (x),\Delta^*(x)\}\Big|_{\sigma
    (x)=M,\Delta(x)= \Delta,\Delta^*(x)=\Delta^*} ,
\end{equation*}
which gives
\begin{eqnarray}
\int d^3x\Omega (M,\Delta,\Delta^*)\,\,&=&\,\,\int d^3x\left
(\frac{M^2}{4G_1}+\frac{\Delta\Delta^*}{4G_2}\right )+\frac
iN\ln\left (
\int\prod_{l=1}^{2}\prod_{b=1}^{N}[d\bar\psi_{lb}][d\psi_{lb}]\exp\Big
(i\int\sum_{k=1}^{2}\Big [\bar\psi_{ka}
D_k\psi_{ka}\right.\nonumber\\&& \left.-
\frac{\Delta^*}{2}\psi_{ka}^TC\psi_{ka}k
-\frac{\Delta}{2}\bar\psi_{ka} C\bar\psi_{ka}^T\Big ] d^3 x \Big
)\right ), \label{9}
\end{eqnarray}
where $D_k=\gamma^\nu i\partial_\nu +\mu_k\gamma^0-M$. To proceed,
let us first point out that without loss of generality the
quantities $\Delta,\Delta^*$ might be considered as real ones.
\footnote{Otherwise,   phases of the complex values
$\Delta,\Delta^*$ might be eliminated by an appropriate
transformation of fermion fields in the path integral (\ref{9}).}
So, in the following we will suppose that
$\Delta=\Delta^*\equiv\Delta$, where $\Delta$ is already a real
quantity. Then, in order to find a convenient expression for the TDP
it is necessary to evaluate the Gaussian path integral in (\ref{9})
(see, e.g., the paper \cite{zhokhov}, where a similar path integral
was calculated). As a result, we obtain the following expression
for the  TDP of the model (1) at zero temperature:
\begin{eqnarray}
\Omega (M,\Delta)=
\frac{M^2}{4G_1}+\frac{\Delta^2}{4G_2}
+i\sum_{k=1}^{2}\int\frac{d^3p}{(2\pi)^3}\ln\Big [(p_0^2-({\cal
E}_{\Delta,k}^+)^2)(p_0^2 -({\cal E}_{\Delta,k}^-)^2)\Big ],
\label{12}
\end{eqnarray}
where $({\cal E}_{\Delta,k}^\pm)^2=E^2+\mu_k^2+\Delta^2\pm 2
\sqrt{M^2\Delta^2+\mu_k^2E^2}$ and $E=\sqrt{M^2+|\vec p|^2}$.
Throughout the paper we suppose that $\mu\ge 0$, $\nu\ge 0$, $M\ge 0$ and
$\Delta\ge 0$. Using in the expression (\ref{12}) a rather general
formula
\begin{eqnarray}
\int_{-\infty}^\infty dp_0\ln\big
(p_0-A)=\mathrm{i}\pi|A|\label{int}
\end{eqnarray}
(obtained rigorously, e.g., in Appendix B of \cite{gubina} and true up to an infinite term independent on real quantity $A$),
it is possible to reduce it to the
following one:
\begin{eqnarray}
\Omega (M,\Delta)\equiv\Omega^{un} (M,\Delta)=
\frac{M^2}{4G_1}+\frac{\Delta^2}{4G_2}
-\sum_{k=1}^{2}\int\frac{d^2p}{(2\pi)^2}\Big ({\cal
E}_{\Delta,k}^++{\cal E}_{\Delta,k}^-\Big ).  \label{13}
\end{eqnarray}
The integral term in (\ref{13}) is an ultraviolet divergent,
hence to obtain any information from this expression we have to
renormalize it.

Note finally that the formulae from this section resemble the
corresponding relations from \cite{zhokhov}. However, there is an
essential difference which is due to the fact that in the present
model we deal with two $O(N)$-multiplets of Dirac fields and,
correspondingly, with two different chemical potentials.

\section{The renormalization procedure }

First of all, let us regularize the zero temperature TDP (\ref{13})
by cutting momenta, i.e. we suppose that $|p_1|<\Lambda$,
$|p_2|<\Lambda$ in (\ref{13}). As a result we have the following
regularized expression (which is finite at finite values of
$\Lambda$):
\begin{eqnarray}
\Omega^{reg} (M,\Delta)=
\frac{M^2}{4G_1}+\frac{\Delta^2}{4G_2}
-\frac{1}{\pi^2}\sum_{k=1}^{2}\int_0^\Lambda dp_1\int_0^\Lambda
dp_2\Big ( {\cal E}_{\Delta,k}^++{\cal E}_{\Delta,k}^-\Big ).
\label{15}
\end{eqnarray}
Let us use in (\ref{15}) the following asymptotic expansion ($k=1,2$)
\begin{eqnarray}
{\cal E}_{\Delta,k}^++{\cal E}_{\Delta,k}^-=2|\vec p|
+\frac{M^2+\Delta^2}{|\vec p|}+{\cal O}(1/|\vec p|^3),\label{16}
\end{eqnarray}
where $|\vec p|=\sqrt{p_1^2+p_2^2}$. (Note, the leading asymptotic terms
in (\ref{16}) do not depend on $\mu_{1,2}$.) Then, upon integration there
term-by-term, it is possible to find
 \begin{eqnarray}
\Omega^{reg}(M,\Delta)&=&M^2\left [\frac
1{4G_1}-\frac{4\Lambda\ln(1+\sqrt{2})}{\pi^2}\right
]\nonumber\\&+&\Delta^2\left [\frac
1{4G_2}-\frac{4\Lambda\ln(1+\sqrt{2})}{\pi^2}\right
]-\frac{4\Lambda^3(\sqrt{2}+\ln(1+\sqrt{2}))}{3\pi^2}+{\cal
O}(\Lambda^0), \label{17}
\end{eqnarray}
where ${\cal O}(\Lambda^0)$ denotes an expression which is finite in
the limit $\Lambda\to \infty$.  Second, we suppose that the bare
coupling constants $G_1$ and $G_2$ depend on the cutoff parameter
$\Lambda$ in such a way that in the limit
$\Lambda\to\infty$ one obtains a finite expression in the square
brackets of (\ref{17}). Clearly, to fulfil this requirement it is
sufficient to require that
 \begin{eqnarray}
\frac 1{4G_1}\equiv \frac
1{4G_1(\Lambda)}=\frac{4\Lambda\ln(1+\sqrt{2})}{\pi^2}+\frac{1}{\pi
g_1}, ~~~\frac 1{4G_2}\equiv \frac
1{4G_2(\Lambda)}=\frac{4\Lambda\ln(1+\sqrt{2})}{\pi^2}+\frac{1}{\pi
g_2}, \label{18}
\end{eqnarray}
where $g_{1,2}$ are finite and $\Lambda$-independent model
parameters with dimensionality of inverse mass. Moreover, since bare
couplings $G_1$ and $G_2$ do not depend on a normalization point,
the same property is also valid for $g_{1,2}$. Hence, taking into
account in (\ref{15}) and (\ref{17}) the relations (\ref{18}) and
ignoring there an infinite $M$- and $\Delta$-independent constant,
one obtains the following {\it renormalized}, i.e. finite,
expression for the TDP
\begin{eqnarray}
\Omega^{ren}(M,\Delta)&=&\lim_{\Lambda\to\infty}
\left\{\Omega^{reg}(M,\Delta)\Big |_{G_1= G_1(\Lambda),G_2=
G_2(\Lambda)}+\frac{4\Lambda^3(\sqrt{2}+\ln(1+\sqrt{2}))}{3\pi^2}\right\}.\label{19}
\end{eqnarray}
It should also be mentioned that the TDP (\ref{19}) is a
renormalization group invariant quantity.

Suppose that $\mu=0$ and $\nu\equiv \mu_B B=0$. In this case
$\mu_{1,2}=0$, so the ${\cal O}(\Lambda^0)$ term in
(\ref{17}) can be calculated explicitly. As a result, we have for the
TDP in this particular case the following expression:
\begin{eqnarray}
V(M,\Delta)\equiv \Omega^{ren}(M,\Delta)\Big |_{\mu=0,\nu=0}=
\frac{M^2}{\pi g_1}+\frac{\Delta^2}{\pi g_2}+\frac{(M+\Delta)^{3}}
{3\pi}+\frac{|M-\Delta|^{3}}{3\pi}.\label{25}
\end{eqnarray}
(The parameters $g_{1,2}$ are introduced in (\ref{18}) in such a way
that at $\mu=\nu=0$ the TDP (\ref{25}) differs by a factor 2 from
the corresponding quantity of the paper \cite{zhokhov}.)

Now, let us obtain an alternative expression for the renormalized
TDP (\ref{19}) at $\mu\ne 0$ and $\nu\ne 0$, i.e. at
$\mu_{1,2}\equiv\mu\pm\nu\ne 0$.  For this purpose one can rewrite
the unrenormalized TDP $\Omega^{un}(M,\Delta)$ (\ref{13}) in the
following way
\begin{eqnarray}
\Omega^{un} (M,\Delta)&=&
\frac{M^2}{4G_1}+\frac{\Delta^2}{4G_2}-\sum_{k=1}^{2}\int\frac{d^2p}{(2\pi)^2}
\left ({\cal E}_{\Delta,k}^+\big |_{\mu,\nu=0}+{\cal E}_{\Delta,k}^-\big |_{\mu,\nu=0}\right )
\nonumber\\
&-&\sum_{k=1}^{2}\int\frac{d^2p}{(2\pi)^2}\Big ({\cal
E}_{\Delta,k}^+ +{\cal E}_{\Delta,k}^--{\cal E}_{\Delta,k}^+\big
|_{\mu,\nu=0}-{\cal E}_{\Delta,k}^-\big |_{\mu,\nu=0}\Big ),
\label{013}
\end{eqnarray}
where for each $k=1,2$
 \begin{eqnarray*}
{\cal E}_{\Delta,k}^+\big |_{\mu,\nu=0}+{\cal E}_{\Delta,k}^-\big |_{\mu,\nu=0}=
\sqrt{|\vec p|^2+(M+\Delta)^2}+\sqrt{|\vec p|^2+(M-\Delta)^2}.
\end{eqnarray*}
Since the leading terms of the asymptotic expansion (\ref{16}) do
not depend on $\mu_{1,2}$, it is clear that the integrals in the
last sum in (\ref{013})  are convergent. Other terms in (\ref{013})
form the unrenormalized TDP of the particular case with $\mu=0$ and
$\nu=0$ which is reduced after renormalization procedure to the
expression (\ref{25}). Hence, after renormalization we obtain from
(\ref{013}) the following finite expression (evidently, it coincides
with renormalized TDP (\ref{19})):
 \begin{eqnarray}
\Omega^{ren} (M,\Delta)=V(M,\Delta)
-\int\frac{d^2p}{(2\pi)^2}\sum_{k=1}^{2}\Big ({\cal
E}_{\Delta,k}^++{\cal E}_{\Delta,k}^-- \sqrt{|\vec
p|^2+(M+\Delta)^2}-\sqrt{|\vec p|^2+(M-\Delta)^2}\Big ), \label{24}
\end{eqnarray}
where $V(M,\Delta)$ is presented in (\ref{25}). The integral terms in
(\ref{24}) can be explicitly calculated. As a result, we have
\begin{eqnarray}
\Omega^{ren}
(M,\Delta)&=&\frac{M^2}{\pi g_1}+\frac{\Delta^2}{\pi g_2}+
\sum_{k=1}^{2}\left\{\frac{1}{6\pi}\left (M+\sqrt{\mu_k^2+\Delta^2}\right )^3+\frac{1}{6\pi}\left |M-\sqrt{\mu_k^2+\Delta^2}\right |^3\right.\nonumber\\
&-&\frac{1}{4\pi}t^+_k\left(M+\sqrt{\mu_k^2+\Delta^2}\right )+
\frac{1}{4\pi}t_k^-\left |M-\sqrt{\mu_k^2+\Delta^2}\right |\nonumber\\
&-&\left.\frac{(\mu_k^2-M^2)\Delta^2}{4\pi |\mu_k|}\ln\left
|\frac{t^+_k+|\mu_k|(M+\sqrt{\mu_k^2+\Delta^2})}{t^-_k+|\mu_k
M-\mu_k\sqrt{\mu_k^2+\Delta^2}|}\right |\right\}, \label{23}
\end{eqnarray}
where $t^\pm_k=M\sqrt{\mu_k^2+\Delta^2}\pm\mu_k^2$. It is not so
evident, but at $\mu_k=0$ ($k=1,2$) the expression (\ref{23}) for
$\Omega^{ren} (M,\Delta)$ coincides with $V(M,\Delta)$ (\ref{25}).

Note also that in the model under consideration we have two
$O(N)$-multiplets of Dirac fields. Due to this reason we have
introduced in (\ref{18}) the parameters $g_{1,2}$ in such a way that
at $\nu=0$ the TDP (\ref{23}) differs by a factor 2 from the
corresponding quantity of the model \cite{zhokhov} with single
multiplet. However, at $\nu\ne 0$, i.e. when $\mu_1\ne\mu_2$, there
is an essential difference between the TDPs of two models.

\section{Phase structure of the model}

As was mentioned above, the coordinates of the global minimum point
$(M_0,\Delta_0)$ of the TDP $\Omega^{ren}(M,\Delta)$ define the ground
state expectation values of auxiliary fields $\sigma (x)$ and
$\Delta (x)$. Namely, $M_0=\vev{\sigma(x)}$ and
$\Delta_0=\vev{\Delta(x)}$. The quantities $M_0$ and $\Delta_0$ are
usually called order parameters, or gaps, because they are
responsible for the phase structure of the model or, in other words,
for the properties of the model ground state (see also the comment
after (\ref{3})). Moreover, the gap $M_0$ is equal to the dynamical
mass of  one-fermionic excitations of the ground state. As a rule,
gaps depend on model parameters as well as on various external
factors. In our consideration the gaps $M_0$ and $\Delta_0$
are certain functions of the free model parameters $g_1$ and $g_2$ and
such external factors as chemical potential $\mu$ and external in-plane magnetic field $B$.
\begin{figure}
 \includegraphics[width=0.45\textwidth]{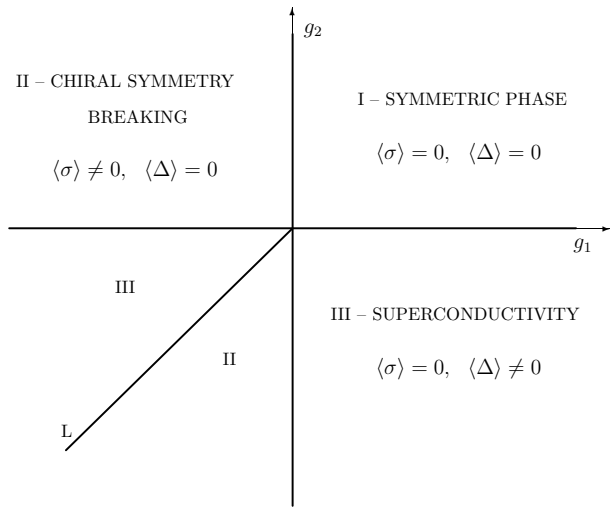}
 \caption{The $(g_1,g_2)$-phase portrait of the model at $\mu=0$ and $B=0$.
 The shorthands I, II and III denote the symmetric, the chiral symmetry
 breaking and the superconducting phases, respectively. In the phase II
 $\vev{\sigma}=-1/g_1$. In the phase III $\vev{\Delta}=-1/g_2$.
On the  curve  L$\equiv\{(g_1,g_2):g_1=g_2\}$, where $g_{1,2}<0$, the
TDP minima corresponding to the phase II and III are equivalent.}
\end{figure}

\subsection{The case $\mu=0$, $B=0$}

First of all, let us discuss the phase structure of the model (1) in
the simplest case when $\mu=0$ and $B=0$. The corresponding TDP is
given in (\ref{25}) by the function $V(M,\Delta)$. Since the global
minimum of this function was already investigated in
\cite{zhokhov,Zhukovsky:2000yd}, although in the framework of
another (2+1)-dimensional GN model, we present at once the phase
structure of the initial model (1) at $\mu=0$ and $B=0$ (see Fig. 1
which is taken from \cite{zhokhov}).

In Fig. 1 the phase portrait of the model is depicted depending on
the values of the free model parameters $g_1$ and $g_2$. There the
plane $(g_1,g_2)$ is  divided into several areas.  In each area one
of the phases I, II or III is implemented. In the phase I, i.e. at
$g_1>0$ and $g_2>0$, the global minimum of the effective potential
$V(M,\Delta)$ is arranged at the origin. So in this case we have
$M_0=\vev{\sigma(x)}=0$ and $\Delta_0=\vev{\Delta(x)}=0$. As a
result, in the phase I both chiral and electromagnetic $U(1)$
symmetries remain intact and fermions are massless. Due to this reason
the phase I is called symmetric. In the phase II, which is allowed
only for $g_1<0$, at the global minimum point $(M_0,\Delta_0)$ the
relations $M_0=-1/g_1$ and
$\Delta_0=0$ are valid. So in this phase chiral symmetry is
spontaneously broken down and fermions acquire dynamically the mass
$M_0$. Finally, in the superconducting phase III, where $g_2<0$, we
have the following values for the gaps $M_0=0$ and
$\Delta_0=-1/g_2$.

Note also that if $g_1=g_2\equiv g$ and, in addition, $g<0$ (it is
just the line L in Fig. 1), then the effective potential (\ref{25})
has two equivalent global minima. The first one, the point
$(M_0=-1/g,\Delta_0=0)$, corresponds to a phase with chiral symmetry
breaking. The second one, i.e. the point $(M_0=0,\Delta_0=-1/g)$,
corresponds to superconductivity.

Clearly, if the cutoff parameter $\Lambda$ is fixed, then the phase
structure of the model can be described in terms of bare coupling
constants $G_1, G_2$ instead of finite quantities $g_1, g_2$.
Indeed, let us first introduce a critical value of the bare couplings,
$G_c=\frac{\pi^2}{16\Lambda\ln(1+\sqrt{2})}$. Then, as it follows
from Fig. 1 and (\ref{18}), at $G_1<G_c$ and $G_2<G_c$ the symmetric
phase I of the model is located. If $G_1>G_c$, $G_2<G_c$ ($G_1<G_c$,
$G_2>G_c$), then the chiral symmetry broken phase II (the
superconducting phase III) is realized. Finally, let us suppose that
both $G_1>G_c$ and $G_2>G_c$. In this case at $G_1>G_2$ ($G_1<G_2$)
we have again the chiral symmetry broken phase II (the
superconducting phase III).
\label{mu0}

\subsection{The case $\mu\ne 0$ and/or $B\ne 0$}

The starting point of our investigations in this case is the TDP
(\ref{23}). The behavior of the global minimum point
$(M_0,\Delta_0)$ of this TDP vs $\mu$ and $B$ supplies us with the phase
structure of the model. Moreover, we are interested in considering
such thermodynamic quantities as particle density $n$, magnetization
$m$ and magnetic susceptibility $\chi$,
\begin{eqnarray}
n=-\frac{\partial\Omega^{ren} (M_0,\Delta_0)}{\partial\mu},~~~~~~
m=-\frac{\partial\Omega^{ren} (M_0,\Delta_0)}{\partial B},~~~~~
\chi=\frac{\partial m}{\partial B}. \label{22}
\end{eqnarray}
In the framework of the model (1) these quantities can be presented in the following form
\begin{eqnarray}
n=n_1+n_2,~~~~~~m=\mu_B\big (n_1-n_2\big ),\label{20}
\end{eqnarray}
where
\begin{eqnarray}
n_1=-\frac{\partial\Omega^{ren}
(M_0,\Delta_0)}{\partial\mu_1},~~~~~~
n_2=-\frac{\partial\Omega^{ren} (M_0,\Delta_0)}{\partial
\mu_2}=-sign(\mu_2)\frac{\partial\Omega^{ren}
(M_0,\Delta_0)}{\partial |\mu_2|}\label{21}
\end{eqnarray}
are densities of particles with spin projection $1/2$ and $-1/2$,
respectively, and $sign(x)$ is the sign-function. Note, in the
particular case with $\mu\ne 0$ and $B=0$ we have
$n_1=n_2$. Therefore, in this case $n\ne 0$, $m= 0$. However, in the
opposite particular case with $\mu=0$ and $B\ne 0$ the relations
$n_1=-n_2$ and, as a result, $n=0$, $m\ne 0$ are valid.
\begin{figure}
 \includegraphics[width=0.45\textwidth]{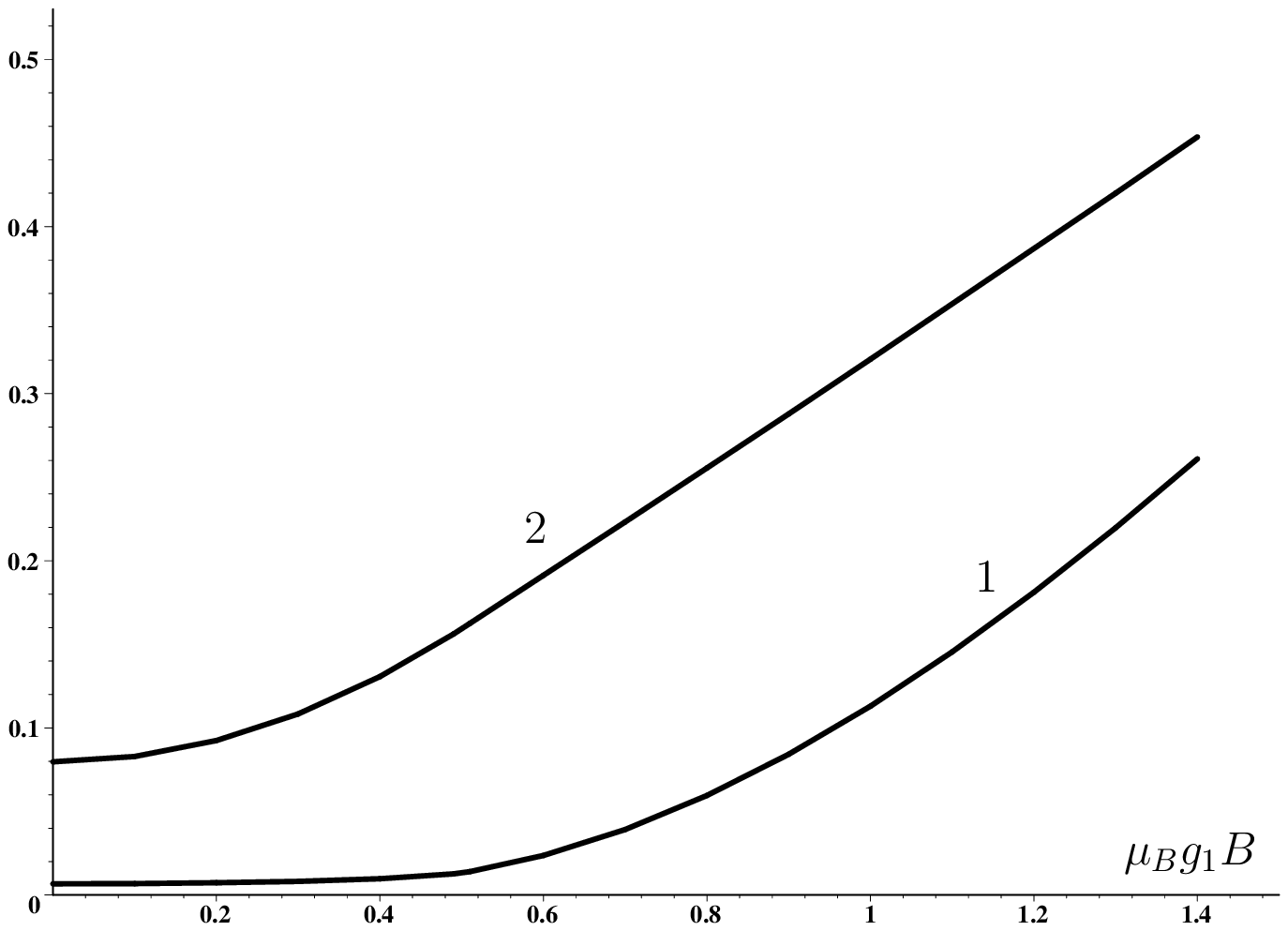}
 \hfill
 \includegraphics[width=0.45\textwidth]{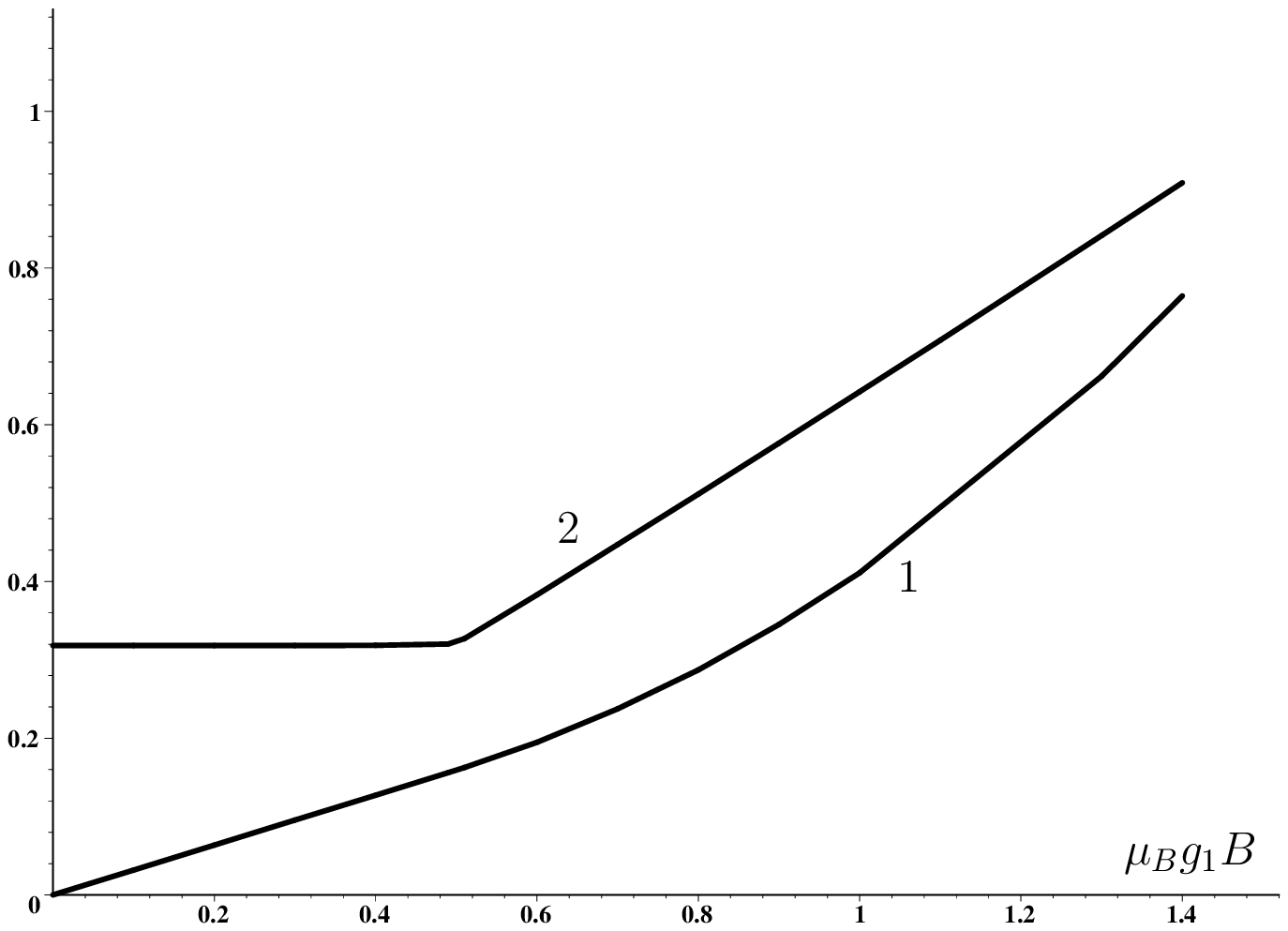}\\
\parbox[t]{0.45\textwidth}{
 \caption{Superconducting gap $\Delta_0$ and particle
density $n$ vs $B$ at arbitrary fixed $g_1>0$ as well as at
$g_2=0.5g_1$ and $\mu=0.5/g_1$. Curves 1 and 2 are the plots of the
dimensionless quantities $g_1\Delta_0$ and $g_1^2n$, respectively. }
 }\hfill
\parbox[t]{0.45\textwidth}{
\caption{Magnetization $m$ and magnetic
susceptibility $\chi$ vs $B$ at arbitrary fixed $g_1>0$ as well as
at $g_2=0.5g_1$ and $\mu=0.5/g_1$. Curves 1 and 2 are the plots of
the dimensionless quantities $g_1^2m/\mu_B$ and $g_1\chi/\mu_B^2$,
respectively. }  }
\end{figure}
It follows from  these formulae that
\begin{eqnarray}
\chi=-\mu_B^2\left [\frac{\partial^2\Omega^{ren}
(M_0,\Delta_0)}{(\partial\mu_1)^2}+\frac{\partial^2\Omega^{ren}
(M_0,\Delta_0)}{(\partial |\mu_2|)^2}\right ].\label{31}
\end{eqnarray}
Numerical and analytical investigations of the
TDP (\ref{23}) show that if $\mu$ and/or $B$ are not zero, then its
global minimum point has one of the
forms $(M_0\ne 0,\Delta_0 =0)$ (the
chiral symmetry breaking phase II) or $(M_0=0,\Delta_0\ne 0)$ (the
superconducting phase III), i.e. the symmetric phase is absent in the
model, if $\mu\ne 0$ and/or $B\ne 0$. Therefore, it is useful to
present the analytical expressions for the thermodynamic quantities
(\ref{22}) in
each of the phases II and III. Namely, for the phase II we have
\begin{eqnarray}
n\big |_{\rm phase~
II}&=&\frac{1}{2\pi}\left [(\mu_1^2-M_0^2)\theta (\mu_1-M_0)+
sign(\mu_2)(\mu_2^2-M_0^2)\theta (|\mu_2|-M_0)\right ],\label{32}\\
m\big |_{\rm phase~
II}&=&\frac{\mu_B}{2\pi}\left [(\mu_1^2-M_0^2)\theta (\mu_1-M_0)-sign(\mu_2)
(\mu_2^2-M_0^2)\theta (|\mu_2|-M_0)\right ],\label{34}\\
\chi\big |_{\rm phase~ II}&=& \frac{\mu_B^2}{\pi}\left [\mu_1\theta
(\mu_1-M_0)+ |\mu_2|\theta (|\mu_2|-M_0)\right ],\label{33}
\end{eqnarray}
whereas for the phase III these quantities look like
\begin{eqnarray}
n\big |_{\rm phase~
III}&=&\frac{1}{2\pi}\left
[\mu_1\sqrt{\mu_1^2+\Delta^2_0}+\Delta^2_0\ln\frac{\mu_1+\sqrt{\mu_1^2+\Delta^2_0}}
{\Delta_0}\right.\nonumber\\
&&\left.~~~~~~~+sign(\mu_2)\Bigg (|\mu_2|\sqrt{\mu_2^2+\Delta^2_0}+\Delta^2_0
\ln\frac{|\mu_2|+\sqrt{\mu_2^2+\Delta^2_0}}{\Delta_0}\Bigg )\right
],\label{35}\\
m\big |_{\rm phase~ III}&=&\frac{\mu_B}{2\pi}\left
[\mu_1\sqrt{\mu_1^2+\Delta^2_0}+\Delta^2_0\ln\frac{\mu_1+\sqrt{\mu_1^2+\Delta^2_0}}{\Delta_0}\right.\nonumber\\
&&\left.~~~~~~~-sign(\mu_2)\Bigg (|\mu_2|\sqrt{\mu_2^2+\Delta^2_0}+\Delta^2_0\ln\frac{|\mu_2|+\sqrt{\mu_2^2+\Delta^2_0}}{\Delta_0}\Bigg )\right
],\label{36}\\
\chi\big |_{\rm phase~ III}&=& \frac{\mu_B^2}{\pi}\left
[\sqrt{\mu_1^2+\Delta^2_0}+\sqrt{\mu_2^2 +\Delta^2_0}\right
],\label{37}\end{eqnarray}

{\bf The case $g_1>0$.} Our investigations show that in this case at
arbitrary nonzero  values of $\mu$ and/or $B$ the superconducting phase
is realized in the system. Since at $g_2<0$ the phenomenon takes place
even at $\mu=0$ and $B=0$, we can say that the chemical potential and/or
in-plane magnetic field {\it enhance} superconductivity, which was
originally generated in this case by a rather strong interaction in
the fermion-fermion channel ($G_2>G_c$). In contrast, at $g_2>0$ the
system is in the symmetric phase if $\mu=0$ and $B=0$. However,
arbitrary small nonzero values of $\mu$ and/or $B$ {\it induce} in this
case the superconductivity. In Fig. 2 and 3 the behavior of the
superconducting gap $\Delta_0$ and such thermodynamic parameters of
the model, as particle density $n$, magnetization $m$ and magnetic
susceptibility $\chi$ vs $B$ are presented at arbitrary fixed
$g_1>0$ and $g_2=0.5g_1$ as well as at fixed chemical potential,
$\mu=0.5/g_1$.\footnote{All the Figs. 2-7 are drawn in terms of dimensionless
quantities which are obtained after multiplication of appropriate
powers of $|g_1|$ with corresponding dimensional quantities. For
example, there instead of $\mu$, $\mu_BB$, $\Delta_0$, $g_2$ we use
their dimensionless analogies $|g_1|\mu$, $\mu_B|g_1|B$,
$|g_1|\Delta_0$, $g_2/|g_1|$. Instead of magnetization $m$ the
dimensionless quantity $g_1^2m/\mu_B$ is depicted there etc. }

\begin{figure}
 \includegraphics[width=0.45\textwidth]{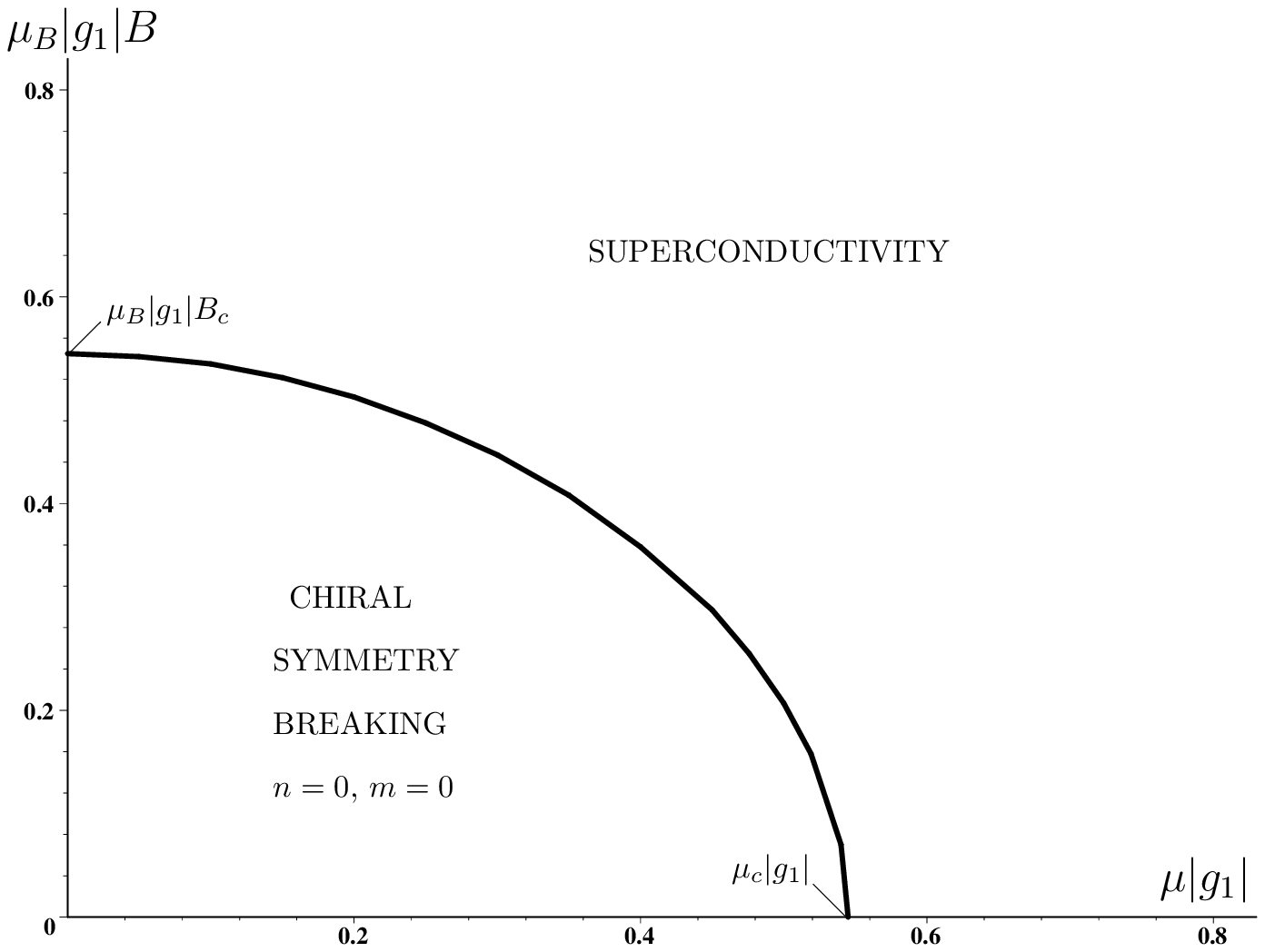}
 \hfill
 \includegraphics[width=0.45\textwidth]{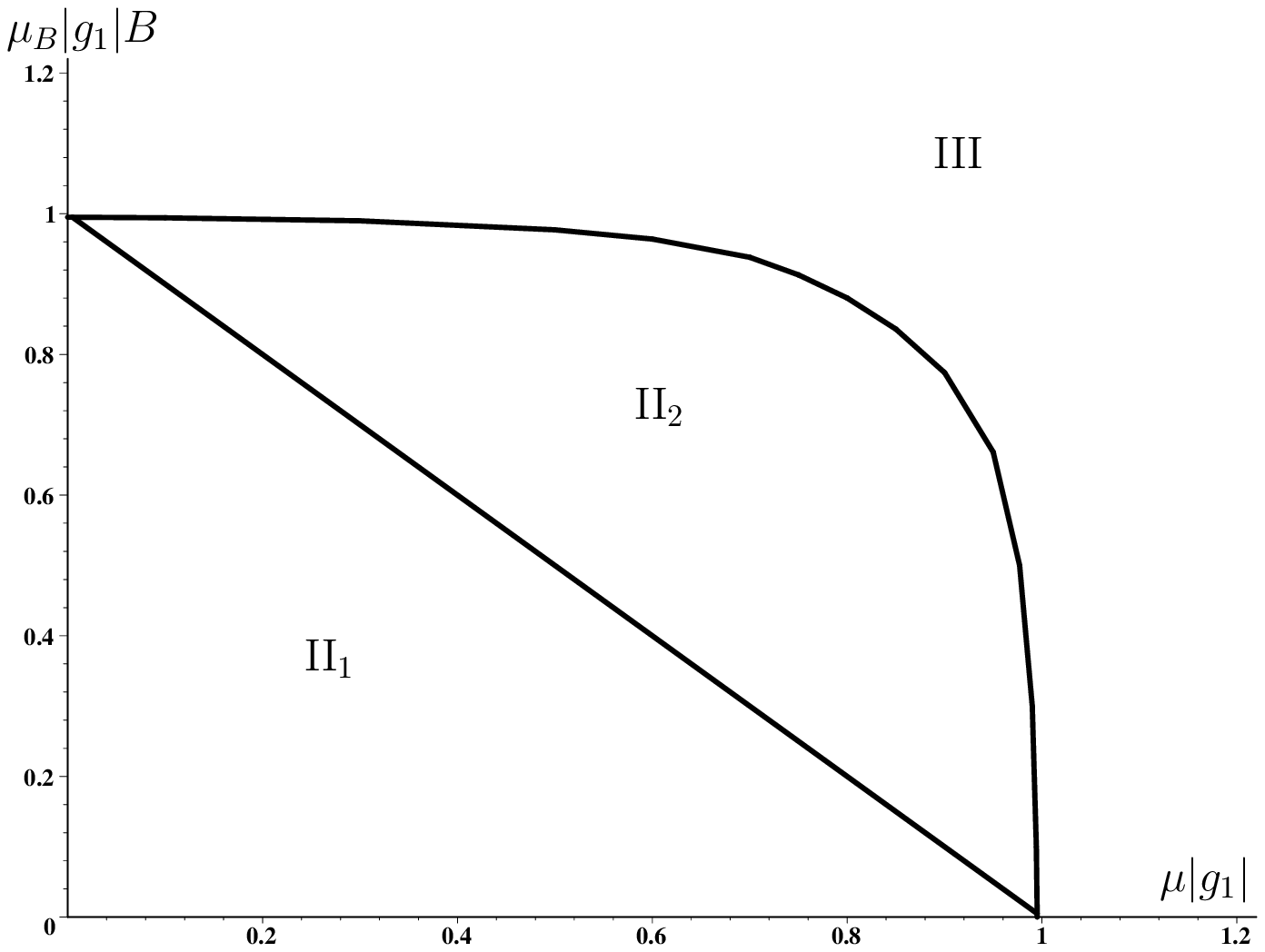}\\
\parbox[t]{0.45\textwidth}{
 \caption{The $(\mu,B)$-phase portrait of the model
 at arbitrary fixed $g_1<0$ as well as at
$g_2=-1.5|g_1|$. }
 }\hfill
\parbox[t]{0.45\textwidth}{
\caption{The $(\mu,B)$-phase portrait of
the model at arbitrary fixed $g_1<0$ as well as at $g_2=0.5|g_1|$.
Here II$_1$ and II$_2$ denote the chiral symmetry breaking phases
with $n=0$, $m=0$ and  $n\ne 0$, $m\ne 0$, respectively. The
notation III stands for the superconducting phase.} }
\end{figure}

{\bf The case $g_1<0$.} We have found that in this case the
$(\mu,B)$-phase structure  of the model is richer than in the case
$g_1>0$. Indeed, supposing that $g_2=-1.5|g_1|$, where $g_1$ is
arbitrary fixed and negative, it is easy to find the $(\mu,B)$-phase
portrait of the model drawn in Fig. 4. There in the chiral symmetry
breaking phase both the particle density $n$ and the magnetization
$m$ are equal to zero. Qualitatively, the similar $(\mu,B)$-phase
structure occurs for each coupling $g_2$ from the interval $g_2\in
(-k|g_1|,-|g_1|)$, where $k\approx 3.08$. However, if $g_2>0$
(recall, $g_1<0$) or $g_2<-k|g_1|$, then the situation is changed
qualitatively. In this case for the representative choice of
coupling constants, $g_2=0.5|g_1|$ ($g_1$ is arbitrary fixed and
negative), the $(\mu,B)$-phase portrait of the model is presented in
Fig. 5. It is clear from this figure that chiral symmetry breaking
phase II is divided into two regions denoted as II$_1$ and II$_2$.
In the region II$_1$ the quantities $n$ and $m$ are still equal to
zero, whereas in the phase II$_2$ both $n\ne 0$ and $m\ne 0$. On the
boundary between chiral symmetry breaking phases II$_1$ and II$_2$
the relation $|g_1|\mu+\mu_B|g_1|B=1$ is valid. Moreover, we
represent in Fig. 6 and 7 the plots of gaps $M_0$ and $\Delta_0$
as well as particle density $n$, magnetization $m$ and magnetic
susceptibility $\chi$ as functions of external in-plane magnetic
field $B$ at $g_2=0.5|g_1|$ and $\mu=0.7/|g_1|$.

Finally, we would like to note that the TDP (\ref{23}) is symmetric
with respect to the transformation $\mu\leftrightarrow\nu$. So, such
physical quantities as $M_0$, $\Delta_0$, $\chi$ remain intact,
whereas $n\leftrightarrow m$ if the transformation
$\mu\leftrightarrow\nu$ is performed. We remark that all figures 2, 3,
6, and 7 show these physical quantities only as functions of
$\nu\equiv \mu_BB$ at some fixed values of $\mu$, i.e. its dependence
on $\mu$ at fixed $\nu$ is not presented in an explicite form in our
paper. However, there is no need for a  special numerical calculations
in this direction, taking into account the above-mentioned symmetry
under the permutation  $\mu\leftrightarrow\nu$. Indeed, let us suppose
that $g_1>0$, $g_2=0.5g_1$ and $\mu_BB=0.5/g_1$.
In this case, in order to consider behavior of the quantities
$\Delta_0$, $n$, $m$ and $\chi$ vs $\mu$, it is sufficient to imagine
that the
horizontal axis in Figs 2 and 3 corresponds to a variable $\mu$. Then
the lines 1 and 2 of  Fig. 9 will mean the plots of $\Delta_0$ and $m$
vs $\mu$, whereas the lines 1 and 2 of Fig. 3 will correspond to the
plots of $n$ and $\chi$ vs $\mu$, respectively. In a similar way it is
possible to extract from Figs 6 and 7 the information about behavior
of
$M_0$, $\Delta_0$, $n$, $m$ and $\chi$ vs $\mu$ in the case of
$g_1<0$, $g_2=0.5|g_1|$ and $\mu_BB=0.7/|g_1|$.

Hence, as it follows from the above consideration (see Figs 4 and 5),
at arbitrary values of $\mu\ge 0$ and
sufficiently strong external in-plane magnetic field
the superconducting (or Cooper pairing) phenomenon appears in
the framework of the (2+1)-dimensional GN-type model (1) (of course, if
$G_2>0$), i.e. an external in-plane magnetic field $B$ promotes
superconductivity for arbitrary relations between coupling constants
$g_{1,2}$ (or, equivalently, $G_{1,2}$). In particular, if $G_{1,2}<G_c$, i.e. $g_{1,2}>0$, then
superconductivity is induced by infinitesimal values of $\mu$ and/or
$\vec B_\parallel$. Moreover, it is clear from our investigations that
superconducting phase of the model is accompanied by a magnetization,
or spin polarization, with positive valued susceptibility. It means
that induced magnetic moment (which disappeares at $\vec
B_\parallel=0$) of the system and external magnetic field $\vec
B_\parallel$ have the same direction, i.e. the superconducting state
is a paramagnetic one. It is not a diamagnetic, as in conventional
superconductivity. Note, the magnetic superconductivity phenomenon
(which includes both the paramagnetic and ferromagnetic
superconductivity) has a long history of investigations in condensed
matter physics (see, e.g., in \cite{umezawa}).
\begin{figure}
 \includegraphics[width=0.45\textwidth]{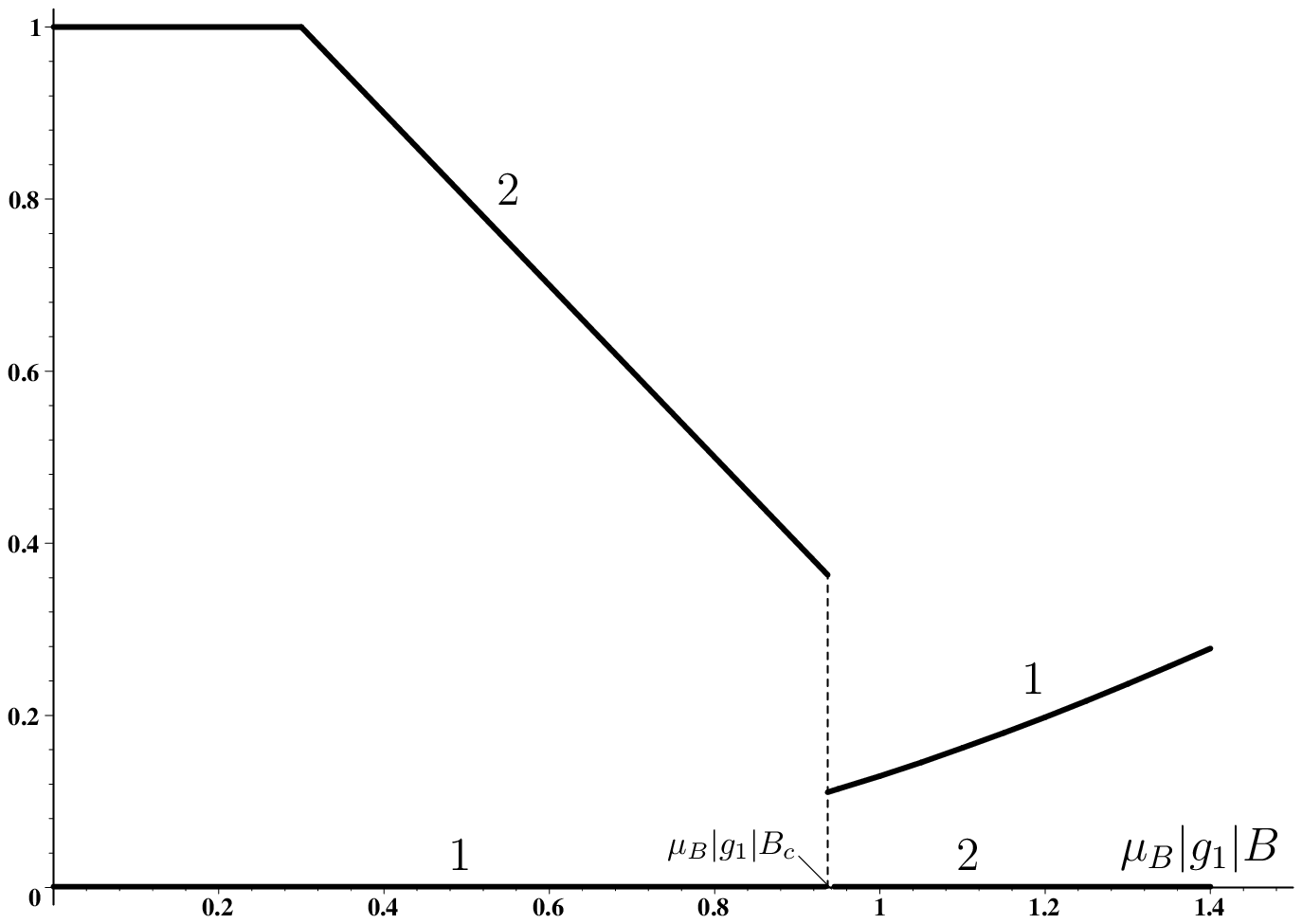}
 \hfill
 \includegraphics[width=0.45\textwidth]{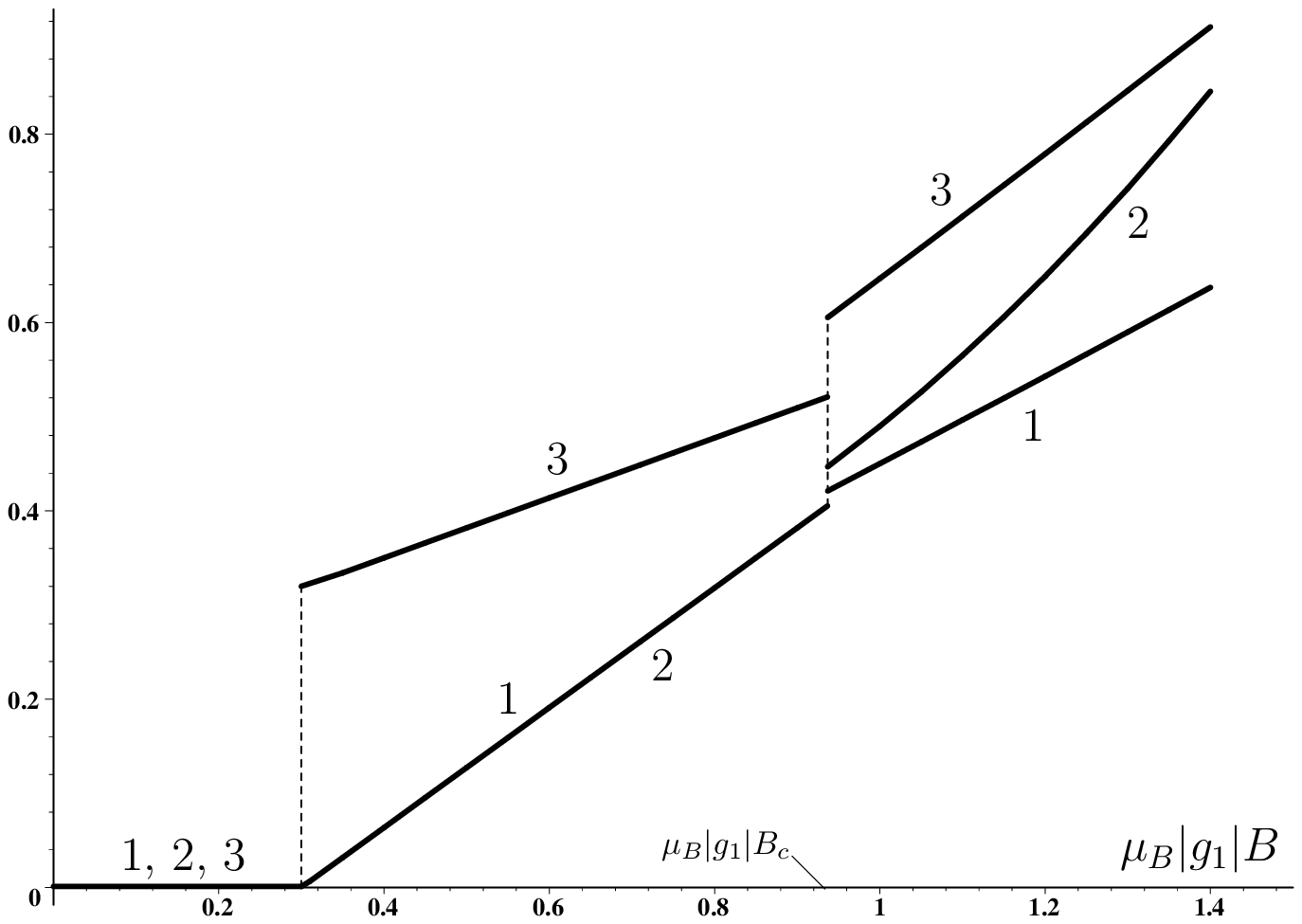}\\
\parbox[t]{0.45\textwidth}{
 \caption{The gaps $M_0$ and $\Delta_0$ vs $B$ at
arbitrary fixed $g_1<0$ as well as at $g_2=0.5|g_1|$ and
$\mu=0.7/|g_1|$. Curves 1 and 2 are the plots of the dimensionless
quantities $|g_1|M_0$ and $|g_1|\Delta_0$, respectively. Here
$\mu_B|g_1|B_c\approx 0.937$. }
 }\hfill
\parbox[t]{0.45\textwidth}{
\caption{Particle density $n$,
magnetization $m$ and magnetic susceptibility $\chi$ vs $B$ at
arbitrary fixed $g_1<0$ as well as at $g_2=0.5|g_1|$ and
$\mu=0.7/|g_1|$. Curves 1, 2 and 3 are the plots of the
dimensionless quantities $g_1^2n$, $g_1^2m/\mu_B$ and
$|g_1|\chi/\mu_B^2$, respectively. Here $\mu_B|g_1|B_c\approx
0.937$. }  }
\end{figure}

\section{Summary and discussion}

In this paper we considered the (2+1)-dimensional GN-type model (1) with
chiral and superconducting  interaction channels under
the influence of an (in-plane) external magnetic field. It is shown
that in-plane magnetic field $\vec B_\parallel$ catalizes a creation
of the paramagnetic superconductivity in the system. It means that if
at $\vec B_\parallel=0$ the electromagnetic $U(1)$ symmetry group was
not broken, then at $|\vec B_\parallel|>B_c$, where a critical field
$B_c$ can be even zero at some particular relations between coupling
constants $G_{1,2}$, there is a spontaneous breaking of $U(1)$. In
addition, if at $\vec B_\parallel=0$ the $U(1)$ symmetry was broken
spontaneously (due to a rather strong interaction in the
fermion-fermion channel), then nonzero values of $\vec B_\parallel$
enhance superconductivity, i.e. the superconducting order parameter
$\Delta_0$ is an increasing function vs $|\vec
B_\parallel|$. Moreover, in-plane magnetic field induces also a
nonzero spin polarization (paramagnetic magnetization) in the
superconducting state.

Various particular cases of the problem have been already considered
earlier. Indeed, twenty years ago, it was found
that at $G_2=0$ and $G_1>0$ an arbitrary weak perpendicular external
magnetic field $\vec B_\perp$ can not only enhance the chiral
symmetry breaking (if $G_1>G_c$) but also induce spontaneous
breaking of the chiral symmetry at $G_1<G_c$ (see, e.g.,
\cite{klimenko3,klimenko11,klimenko12,klimenko13,Gusynin}). In contrast, recently it was found
\cite{caldas} that an application of an external parallel to the
system plane (in-plane) magnetic field $\vec B_\parallel$ results in
the restoration of chiral symmetry (recall, in this case $G_2=0$).
For an explanation of such a different reaction of the chiral
symmetry on perpendicular and parallel magnetic fields one should
remember that $\vec B_\perp$ acts on the orbital angular momentum of
electrons, whereas the in-plane field $\vec B_\parallel$ couples
only to their spin. In the last case, due to the Zeeman effect,
increase of the electron chemical potential takes place which
eventually leads to the restoration of the chiral symmetry
\cite{klimenko2,klimenko3,inagaki,kanemura,klimenk}.

On the contrary, the results of our present paper demonstrate that
the response of the electromagnetic $U(1)$ symmetry
of (2+1)-dimensional GN-type models to the applied external magnetic
field is completely different. Indeed, it was shown, in
the framework of the (2+1)-dimensional GN-type model with nonzero
coupling $G_2$ of the superconducting channel and at $G_1=0$, that
restoration of the $U(1)$ symmetry takes place at sufficiently strong
external perpendicular field $\vec B_\perp$ \cite{marino2}.
Moreover, the main result of our paper is that if only an external
in-plane magnetic field is included in the
(2+1)-dimensional GN model (1) (at least with $G_2>0$), then at
sufficiently strong values of $\vec B_\parallel$ the
superconductivity phenomenon appears, i.e. the spontaneous breaking
of the electromagnetic $U(1)$ symmetry is originated. In particular,
if an interaction in both channels of the GN model (1) is
sufficiently weak, i.e. $G_{1,2}<G_c$ (or $g_{1,2}>0$), then
superconductivity is induced by infinitesimal values of $\vec
B_\parallel$.

To understand the last property, one should take into account that due
to the Zeeman effect, i.e. due to the Zeeman $\nu$-terms in Lagrangian
(1), an external in-plane magnetic field $\vec B_\parallel$ induces
(at $\mu=0$)
Fermi surface for electrons in any planar system. However, in our
model,  since $G_2>0$, there exists an attractive interaction between
electrons just above the Fermi surface. So, the Fermi surface is
unstable in favor of Bose--Einstein condensate of Cooper pairs, and a
new (superconducting) ground state is formed in the system.

Finally, we would like to recall
some QFT examples in which
similar effects, although in a quite different physical contexts, are
also observed. First of all, it is important to note that an ability
of strong external magnetic field to induce electromagnetic
superconductivity of vacuum was recently established   in the
framework of quantum chromodynamics (see, e.g., the review
\cite{Chernodub:2012tf}). Next, it is argued in some papers (see,
e.g., \cite{tatsumi})
that ferromagnetism and superconductivity might coexist in dense quark
matter, thus explaining super strong magnetic fields of
magnetars.\footnote{
In ferromagnetic superconductors the spin polarization, or
magnetisation, is originated spontaneously, i.e. in this case there is
a spontaneous breaking of the rotational invariance. In contrast, in
our model (1) the magnetization of the superconducting state exists
only in the presence of an external magnetic field $\vec B_\parallel$,
which destroys the rotational symmetry of the model explicitly.}

\end{document}